\documentclass[prd,twocolumn,tightenlines,showpacs,nofootinbib,superscriptaddress,nobalancelastpage]{revtex4-1} 

\usepackage{graphics}     
\usepackage{graphicx}     
\usepackage{longtable}     
\usepackage{url}          
\usepackage{bm}           
\usepackage{natbib}
\usepackage{textpos}
\usepackage{amsfonts,amsmath,amssymb,mathrsfs,bm,amsthm}
\usepackage{float}
\usepackage{booktabs}
\usepackage{array}
\usepackage{tabu}
\usepackage{dcolumn}
\usepackage{rotating}
\usepackage{ulem}
\usepackage{soul}

\newcommand{\RN}[1]{%
  \textup{\uppercase\expandafter{\romannumeral#1}}%
}

\usepackage{nicefrac}
\usepackage{amsthm}
\usepackage{color}
\usepackage{cancel}
\usepackage{appendix}
\usepackage{makecell}

\usepackage[colorlinks=true,linkcolor=black,citecolor=blue,urlcolor=blue,bookmarksopen]{hyperref}

\newcommand{\appropto}{\mathrel{\vcenter{
  \offinterlineskip\halign{\hfil$##$\cr
    \propto\cr\noalign{\kern2pt}\sim\cr\noalign{\kern-2pt}}}}}
\renewcommand{\v}[1]{\boldsymbol{#1}}		

\begin{document}
\title{Revisiting spin-dependent forces mediated by new bosons: Potentials in the coordinate-space representation for macroscopic- and atomic-scale experiments}

\date{\today}

\author{Pavel Fadeev}
\affiliation{Helmholtz Institute Mainz, Johannes Gutenberg University, 55099 Mainz, Germany}

\author{Yevgeny V.~Stadnik}
\affiliation{Helmholtz Institute Mainz, Johannes Gutenberg University, 55099 Mainz, Germany}

\author{Filip Ficek}
\affiliation{Institute of Physics, Jagiellonian University, \L{}ojasiewicza 11, 30-348 Krak\'ow, Poland}

\author{Mikhail G.~Kozlov}
\affiliation{Petersburg Nuclear Physics Institute of NRC ``Kurchatov Institute'', Gatchina 188300, Russia}
\affiliation{St.~Petersburg Electrotechnical University LETI, Prof.~Popov Str.~5, 197376 St.~Petersburg, Russia}

\author{Victor V.~Flambaum}
\affiliation{Helmholtz Institute Mainz, Johannes Gutenberg University, 55099 Mainz, Germany}
\affiliation{School of Physics, University of New South Wales, Sydney, New South Wales 2052, Australia}

\author{Dmitry Budker}
\affiliation{Helmholtz Institute Mainz, Johannes Gutenberg University, 55099 Mainz, Germany}
\affiliation{Department of Physics, University of California at Berkeley, Berkeley, California 94720-7300, USA}
\affiliation{Nuclear Science Division, Lawrence Berkeley National Laboratory, Berkeley, California 94720, USA}

\begin{abstract}
The exchange of spin-0 or spin-1 bosons between fermions or spin-polarised macroscopic objects gives rise to various spin-dependent potentials. 
We derive the coordinate-space non-relativistic potentials induced by the exchange of such bosons, including contact terms that can play an important role in atomic-scale phenomena, and correct for errors and omissions in the literature. 
We summarise the properties of the potentials and their relevance for various types of experiments. 
These potentials underpin the interpretation of experiments that search for new bosons, including spectroscopy, torsion-pendulum measurements, magnetometry, parity nonconservation and electric dipole moment experiments. 
\end{abstract}


\maketitle

\section{Introduction}
\label{Sec:Intro}
There are four known types of interactions in nature --- electromagnetic, strong, weak and gravitational. 
Still, additional interactions may exist. 
For example, the exchange of a new spin-0 or spin-1 boson between two fermions produces a plethora of ``exotic'' interaction potentials \cite{Dob}. 
Yet-to-be-discovered bosons may solve several outstanding puzzles. 
The axion (a spin-0 boson) may explain the apparent absence of \textit{CP} violation in strong interactions \cite{Ax1,Ax2,Ax3,Ax4,Ax5,Ax6,Pec1,Pec2}. 
The observed dark matter \cite{Ber} and dark energy \cite{Frieman} may also be explained by the existence of new bosonic particles. 
The possibility to solve such central questions motivates numerous searches for new bosons. 
Recent examples of searches for new forces mediated by such bosons can be found in Refs.~\cite{Ni1999,Adelberger2006,Baessler2007,Hammond2007,Adelberger2008,Romalis2009,Serebrov2009,Ignatovich2009,Serebrov2010,Petukhov2010,Hoedl2011,Raffelt2012,Snow2013,Tullney2013,Chu2013,Bulatowicz2013,Musolf2014,ARIADNE,Stadnik2015NMBE,Afach2015,Heckel2015,Leefer2016,Ficek2017,Ruoso2017,Stadnik2017vector,Stadnik2017axion,Rong2017,RMP2017,Ji,Kim,Leslie,Hunter2013,Karshenboim2010A,Karshenboim2010B,Karshenboim2011,Korobov2014,Soreq2017,Fadeev2018,Stadnik2018axion,Hunter2014,Rong2018}. 

In Ref.~\cite{Moody} (see also the earlier papers \cite{Bouchiat1975EDM,Anselm1982}), the three distinct non-relativistic potentials arising from the exchange of a spin-0 boson between spin-polarised and spin-unpolarised bodies were presented. 
Later, Ref.~\cite{Dob} expanded this list to include additional long-range non-relativistic potentials arising from the exchange of spin-0 bosons and spin-1 bosons (such as $Z'$ bosons and paraphotons). 
These potentials were presented in Ref.~\cite{Dob} in a mixed momentum- and coordinate-space representation, which is convenient when the relative velocity between two bodies can be described by a classical vector, such as in the macroscopic-scale experiments of Refs.~\cite{Ji,Kim,Leslie,Hunter2013}. 
However, in phenomena that arise on the (sub)atomic scale, the relative velocity between two particles can no longer be described by a classical vector, but must instead be described by a quantum vector operator (see, for example, \cite{Ficek2017,Stadnik2017vector,Stadnik2018axion}). 

Furthermore, the potentials induced by the exchange of bosons in general contain not only long-range terms, but also short-range (contact) terms, which can play an important role in atomic-scale experiments. 
For example, the usual magnetic dipole-dipole interaction between atomic electrons and the nucleus (mediated by the exchange of photons) contains both long-range and contact terms. 
For atomic states with zero electron orbital angular momentum (which are described by spherically symmetric wavefunctions), the expectation value of the long-range part of the magnetic dipole-dipole interaction vanishes and the entire contribution to the hyperfine energy shift comes from the contact part of the magnetic interaction \cite{LL4}. 

In the present paper, we derive the coordinate-space non-relativistic potentials, including contact terms. 
These potentials are particularly important in searches for new spin-dependent forces based on atomic-scale experiments (such as in \cite{Ficek2017,Fadeev2018}) and on macroscopic-scale experiments \cite{Adelberger2006,Adelberger2008,Ji,Kim,Leslie,Hunter2013}.
In atomic systems that satisfy $Z\alpha \ll 1$, where $Z$ is the nuclear charge and $\alpha \approx 1/137$ is the fine-structure constant at zero momentum transfer, the velocities of the particles are small. 
For example, in atomic hydrogen, the expectation value of the square of the velocity of the electron orbiting the nucleus (in natural relativistic units, $\hbar = c = 1$) is
$\left< v^2 \right> \sim \alpha^2 \sim 10^{-4}$. 
In macroscopic-scale experiments that search for velocity-dependent effects due to the relative motion of Earth and the Sun, the square of the relative velocity is $v^2 \sim 10^{-8}$. 

The structure of this paper is as follows. 
In Sec.~\ref{Sec:Potentials}, we derive the coordinate-space non-relativistic potentials induced by the exchange of spin-0 and spin-1 bosons. 
In Sec.~\ref{Sec:Discussion}, we discuss the properties and nuances of these potentials, and point out several erroneous results and omissions in the earlier literature. Some of the more technical details are presented in the Appendices.


 \begin{figure}
\includegraphics[width=0.2\textwidth]{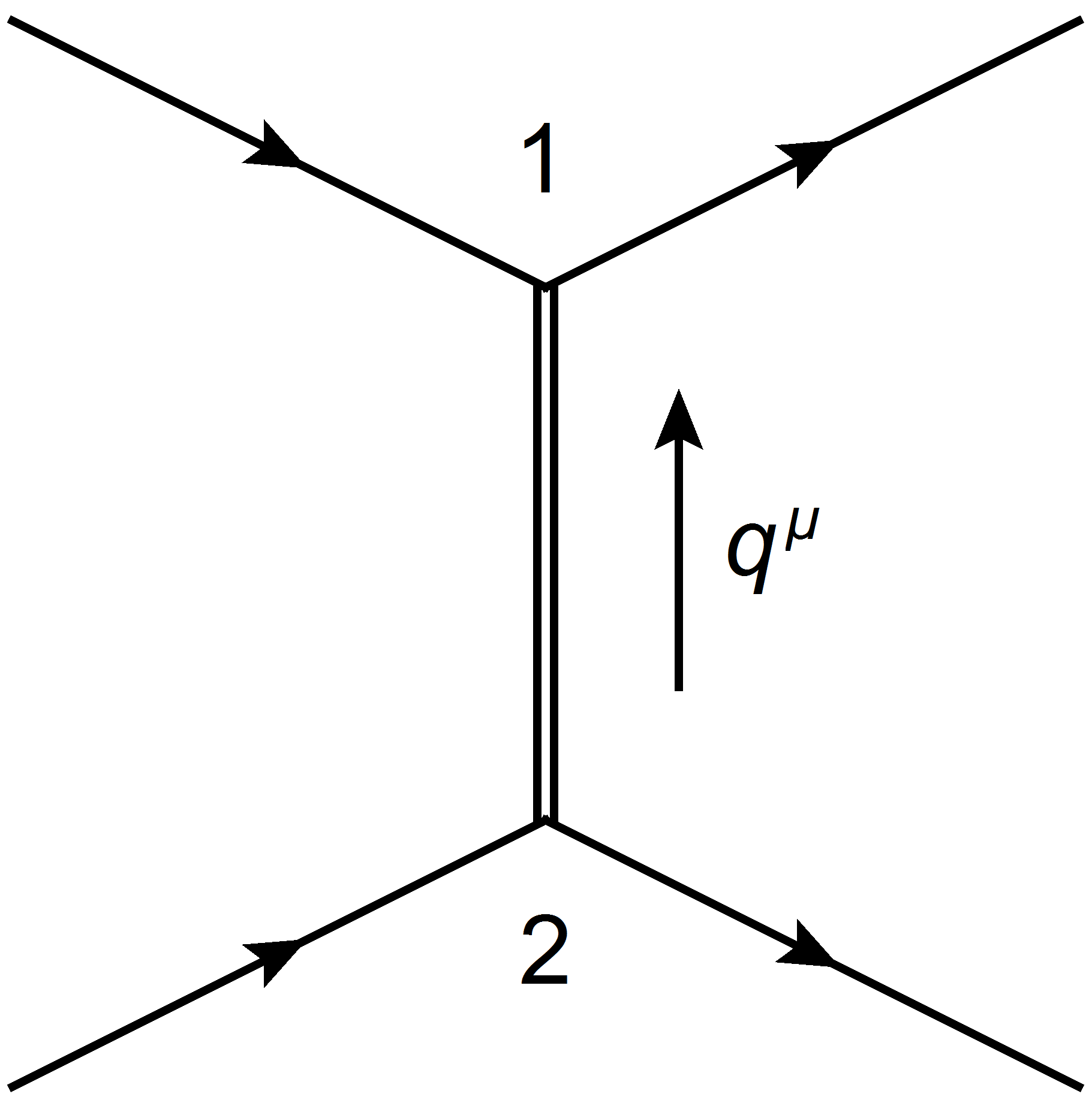}
\caption{Elastic scattering of two fermions with masses $m_1$ and $m_2$ and spins $\v{s}_1$ and $\v{s}_2$, respectively, mediated by a boson of mass $M$ with four-momentum $q^\mu$ that is transferred from fermion 2 to fermion 1. } 
\label{fig:Scattering}
\end{figure}

\section{Coordinate-space potentials}
\label{Sec:Potentials}

Consider the elastic scattering of two fermions with masses $m_1$ and $m_2$ and spins $\v{s}_1$ and $\v{s}_2$, respectively, mediated by a boson of mass $M$ with four-momentum $q^\mu$ that is transferred from fermion 2 to fermion 1 (Fig.~\ref{fig:Scattering}). 
We focus on three types of bosons in the present work --- a spin-0 boson $\phi$ (which can be either massive or massless), a massive spin-1 boson $Z'$ and a massless spin-1 boson $\gamma '$. 
Each boson has its own set of local Lorentz-invariant interactions with the standard-model fermions $\psi$ \cite{Dob,Dobrescu2005}:
\begin{equation} 
 \label{Lag3}
\mathcal{L}_{\phi} = \phi \sum_\psi \bar{\psi}  \left( g^s_\psi + i \gamma_{5} g^p_\psi \right) \psi  \, ,
\end{equation}
\begin{equation} 
 \label{Lag2}
\mathcal{L}_{Z'} = Z'_{\mu} \sum_\psi \bar{\psi}\gamma^{\mu}\left(g^V_\psi + \gamma_{5}  g^A_\psi  \right) \psi  \, ,
\end{equation}
\begin{equation} 
 \label{Lag1}
\mathcal{L}_{\gamma '} = \frac{v_{h}}{\Lambda^{2}}P_{\mu\nu} \sum_\psi  \bar{\psi}\sigma^{\mu\nu} \left[ \mathrm{Re} (C_{\psi}) + i \gamma_{5} \mathrm{Im} (C_{\psi})  \right] \psi   \, . 
\end{equation}
Here $\psi$ denotes the fermion field (for instance, $\psi = e$ for an electron, and $\psi = N$ for a nucleon), $P_{\mu\nu} = \partial_{\mu} A_{\nu} - \partial_{\nu} A_{\mu}$ is the field strength tensor of the massless paraphoton field $A_{\mu}$, $\sigma^{\mu\nu} = \frac{i}{2} [\gamma^\mu , \gamma^\nu]$, and $\gamma^\mu$, $\gamma_5 = i \gamma^0 \gamma^1 \gamma^2 \gamma^3$ are Dirac matrices. 
The dimensionless interaction constants $g^s_\psi$, $g^p_\psi$, $g^V_\psi$, $g^A_\psi$, $\mathrm{Re} (C_{\psi})$, $\mathrm{Im} (C_{\psi})$ parametrise the scalar, pseudoscalar, vector, pseudovector, tensor and pseudotensor interaction strengths, respectively. 
The Higgs vacuum expectation value is denoted by $v_h$, and $\Lambda$ is the ultraviolet energy cutoff scale for Lagrangian~(\ref{Lag1}). 

We have chosen the interactions in Eqs.~(\ref{Lag3}) -- (\ref{Lag1}), since this set of interactions spans the full space of Lorentz-invariant Dirac operators. 
The case of a massive spin-1 boson is distinguished from the case of a massless spin-1 boson by the presence of a longitudinal polarisation, and so we treat the massless and massive cases separately. 

We derive the coordinate-space non-relativistic potentials associated with the interactions in Eqs.~(\ref{Lag3}) -- (\ref{Lag1}) by applying the Feynman diagrammatic technique, which is described in detail in standard textbooks \cite{Sakurai1967,Gross1999}. 
We summarise our conventions, along with useful identities and Fourier transforms in Appendices~\ref{AppA} -- \ref{AppC}. 
We present the detailed derivations of three potentials in Appendix~\ref{AppD}. 

Each of the Lagrangians in Eqs.~(\ref{Lag3}) -- (\ref{Lag1}) contains a sum of two terms, which correspond to two types of vertices. There are three distinct combinations of these two vertices for the scattering of two fermions, and so each Lagrangian can give rise to three distinct potentials. In total, the following nine non-relativistic potentials result:
\begin{widetext}
\begin{equation}
\label{scalar-scalar_potential}
V_{ss}(\v{r}) =  -  g_1^s g_2^s \underbrace{ \frac{e^{-M r}}{ 4 \pi r} }_{\mathcal{V}_1} \, ,
\end{equation}
\begin{equation}
\label{scalar-pseudoscalar_potential}
V_{ps}(\v{r}) = - g_1^p g_2^s \underbrace{ \v{\sigma}_1 \cdot \hat{\v{r}} \left( \frac{1}{r^2} + \frac{M}{r} \right) \frac{e^{-M r}}{8 \pi m_1} }_{\mathcal{V}_{9,10}} \, ,
\end{equation}
\begin{equation}
\label{pseudoscalar-pseudoscalar_potential}
V_{pp}(\v{r}) = -  \frac{g_1^p g_2^p}{4} \underbrace{ \left[  \v{\sigma}_1 \cdot \v{\sigma }_2 \left[ \frac{1}{r^3} + \frac{M}{r^2} + \frac{4 \pi}{3} \delta(\v{r}) \right]
- \left( \v{\sigma}_1 \cdot \hat{\v{r}} \right) \left( \v{\sigma }_2 \cdot \hat{\v{r}} \right)  \left[ \frac{3}{r^3} + \frac{3M}{r^2} + \frac{M^2}{r} \right]   \right] \frac{e^{-M r}}{4 \pi m_1 m_2} }_{\mathcal{V}_3} \, ,
\end{equation}
\begin{equation}
\label{vector-vector_potential}
V_{VV}(\v{r}) = g_1^V g_2^V \underbrace{ \frac{e^{-M r}}{4 \pi r} }_{\mathcal{V}_1}
+  \frac{g_1^V g_2^V}{4} \underbrace{ \left[ \v{\sigma}_1 \cdot \v{\sigma }_2  \left[ \frac{1}{r^3} + \frac{M}{r^2} + \frac{M^2}{r} - \frac{8 \pi}{3} \delta(\v{r}) \right]  -  \left( \v{\sigma}_1 \cdot \hat{\v{r}} \right) \left( \v{\sigma }_2 \cdot \hat{\v{r}} \right)  \left[ \frac{3}{r^3} + \frac{3M}{r^2} + \frac{M^2}{r} \right]  \right] \frac{e^{-M r}}{4 \pi m_1 m_2} }_{\mathcal{V}_2 + \mathcal{V}_3} \, ,
\end{equation}
\begin{equation}
\label{pseudovector-vector_potential}
V_{AV}(\v{r}) =  g_1^A g_2^V \underbrace{ \v{\sigma}_1 \cdot \left\{ \frac{ \v{p}_1}{m_1} - \frac{ \v{p}_2}{m_2} , \frac{e^{-M r}}{ 8 \pi r} \right\} }_{\mathcal{V}_{12,13}} 
-  \frac{g_1^A g_2^V}{2}  \underbrace{ \left(\v{\sigma}_1 \times \v{\sigma }_2 \right) \cdot \hat{\v{r}}  \left( \frac{1}{r^2} + \frac{M}{r} \right)  \frac{e^{-M r}}{4 \pi m_2} }_{\mathcal{V}_{11}} \, ,
\end{equation}
\begin{equation}
\label{pseudovector-pseudovector_potential}
V_{AA}(\v{r}) =  - g_1^A g_2^A \underbrace{ \v{\sigma}_1 \cdot \v{\sigma }_2 \frac{e^{-M r}}{4 \pi r} }_{\mathcal{V}_2} 
-  \frac{g_1^A g_2^A m_1 m_2}{ M^2} \underbrace{ \left[ \v{\sigma}_1 \cdot \v{\sigma }_2 \left[ \frac{1}{r^3} + \frac{M}{r^2} + \frac{4 \pi}{3} \delta(\v{r}) \right]  -  \left( \v{\sigma}_1 \cdot \hat{\v{r}} \right) \left( \v{\sigma }_2 \cdot \hat{\v{r}} \right)  \left[ \frac{3}{r^3} + \frac{3M}{r^2} + \frac{M^2}{r} \right]  \right] \frac{e^{-M r}}{4 \pi m_1 m_2} }_{\mathcal{V}_3} \, ,
\end{equation}
\begin{equation}
\label{tensor-tensor_potential}
V_{TT}(\v{r}) =  \frac{4 v_h^2 \mathrm{Re}(C_1) \mathrm{Re}(C_2) m_1 m_2}{\Lambda^4 } 
\underbrace{ \left[  \v{\sigma}_1 \cdot \v{\sigma }_2  \left[ \frac{1}{r^3} - \frac{8 \pi}{3} \delta(\v{r}) \right]  - \left( \v{\sigma}_1 \cdot \hat{\v{r}} \right) \left( \v{\sigma }_2 \cdot \hat{\v{r}} \right)  \frac{3}{r^3}  \right] \frac{1}{4 \pi m_1 m_2}  }_{\mathcal{V}_2 + \mathcal{V}_3} \, ,
\end{equation}
\begin{align}
\label{pseudotensor-tensor_potential}
V_{\tilde{T}T}(\v{r}) = &  \frac{4 v_h^2 \mathrm{Im}(C_1) \mathrm{Re}(C_2) m_1 m_2 }{ \Lambda^4 } \underbrace{ \left[ \left(\v{\sigma}_1 \times \v{\sigma}_2 \right) \cdot \left\{ \frac{\v{p}_1}{m_1} - \frac{\v{p}_2}{m_2}  , ~ 
 \frac{1}{r^3} + \frac{4 \pi}{3} \delta(\v{r}) \right\} \right] \frac{1}{8 \pi m_1 m_2} }_{  \mathcal{V}_{14}  }  \notag \\
&+  \frac{4 v_h^2 \mathrm{Im}(C_1) \mathrm{Re}(C_2) m_1 m_2}{ \Lambda^4 } \underbrace{ \left\{   \left( \frac{\v{p}_1}{m_1} - \frac{\v{p}_2}{m_2} \right)_i  , ~ \frac{3 \left(\v{\sigma}_1 \cdot \hat{\v{r}} \right) \left(\v{\sigma}_2 \times \hat{\v{r}} \right)_i }{8 \pi m_1 m_2 r^3} \right\} }_{\mathcal{V}_{15}} 
-  \frac{2 v_h^2 \mathrm{Im}(C_1) \mathrm{Re}(C_2) m_1 m_2 }{\Lambda^4} \underbrace{ \frac{\v{\sigma}_1 \cdot [\v{\nabla} \delta(\v{r})] }{ m_1 m_2^2 }  }_{ \mathcal{V}_{9,10} }   \, , 
\end{align}
\begin{equation}
\label{pseudotensor-pseudotensor_potential}
V_{\tilde{T}\tilde{T}}(\v{r}) =  \frac{4 v_h^2 \mathrm{Im}(C_1) \mathrm{Im}(C_2) m_1 m_2}{\Lambda^4 } 
\underbrace{ \left[  \v{\sigma}_1 \cdot \v{\sigma }_2 \left[ \frac{1}{r^3} + \frac{4 \pi}{3} \delta(\v{r}) \right]  -  \left( \v{\sigma}_1 \cdot \hat{\v{r}} \right) \left( \v{\sigma }_2 \cdot \hat{\v{r}} \right)   \frac{3}{r^3}  \right]  \frac{1}{4 \pi m_1 m_2}  }_{\mathcal{V}_3} \, . 
\end{equation}
\end{widetext}
In these expressions, $\v{\sigma}_1$ and $\v{\sigma}_2$ denote the Pauli spin-matrix vectors of the two fermions, $\hat{\v{r}}$ is the unit vector directed from fermion 2 to fermion 1, $r$ is the distance between the two fermions, and $\{A,B\} \equiv AB + BA$ is the anticommutator of two operators $A$ and $B$. 
The momenta $\v{p}_1= -i\v{\nabla}_1$ and $\v{p}_2=-i\v{\nabla}_2$ are vector differential operators in coordinate space. 

For the sake of comparison with previous literature, we matched the individual terms in Eqs.~(\ref{scalar-scalar_potential}) -- (\ref{pseudotensor-pseudotensor_potential}) onto the $\mathcal{V}_i$ potential terms of Ref.~\cite{Dob}.  
In the momentum-space representation, the spin-momentum structures of $\mathcal{V}_2$ in Eq.~(\ref{vector-vector_potential}), $\mathcal{V}_{9,10}$ in Eq.~(\ref{pseudotensor-tensor_potential}), and  $\mathcal{V}_{14}$ in Eq.~(\ref{pseudotensor-tensor_potential}) are multiplied by $\v{q}^2$, the square of the spatial components of the transferred momentum. 
Thus, although $\mathcal{V}_{9,10}$ has different forms in Eqs.~(\ref{scalar-pseudoscalar_potential}) and (\ref{pseudotensor-tensor_potential}), the underlying spin-momentum structure of these potential terms in the momentum-space representation is the same up to a factor of $\v{q}^2$. 

The potentials in Eqs.~(\ref{scalar-pseudoscalar_potential}), (\ref{pseudovector-vector_potential}), and (\ref{pseudotensor-tensor_potential}) are written in an abbreviated form lacking symmetry under the permutation of particle indices $1 \leftrightarrow 2$. In the case of these potentials, we must add the terms obtained by permuting the particle indices $1 \leftrightarrow 2$. 


Once the above terms are added, the potentials in Eqs.~(\ref{scalar-scalar_potential}) -- (\ref{pseudotensor-pseudotensor_potential}) are symmetric with respect to the permutation of particle indices $1 \leftrightarrow 2$, and hence do not vanish for identical particles. 
In order to determine whether or not specific matrix elements of the potentials in Eqs.~(\ref{scalar-scalar_potential}) -- (\ref{pseudotensor-pseudotensor_potential}) vanish for identical particles, one needs to take into account the fact that the overall wavefunction of indistinguishable fermions is antisymmetric under permutation of fermions.

We note that Eqs.~(\ref{scalar-scalar_potential}) -- (\ref{pseudotensor-pseudotensor_potential}) contain fewer than the 16 terms presented in Ref.~\cite{Dob}, since we are interested in the non-relativistic limit. 
For example, the $\mathcal{V}_8$ term of Ref.~\cite{Dob} arises as an $\mathcal{O}(v^2)$ relativistic correction to the $\mathcal{V}_2$ term in Eq.~(\ref{pseudovector-pseudovector_potential}) and hence can be neglected in non-relativistic systems. 

We have chosen to sort the potentials in Eqs.~(\ref{scalar-scalar_potential}) -- (\ref{pseudotensor-pseudotensor_potential}) by their types of physical couplings, in contrast to sorting them into 16 groups by their mathematical spin-momentum structure as in \cite{Dob}. 
We believe that our classification is more useful from a physicist's point of view, since one is ultimately interested in the physical coupling constants of a particular model. 

Other representations of these coordinate-space potentials are possible. 
In Appendix~\ref{AppE}, we present these potentials in a semi-relativistic form that is convenient for numerical atomic calculations using Dirac-Hartree-Fock wavefunctions. 

\section{Discussion}
\label{Sec:Discussion}
We first point out a number of erroneous results and omissions in the earlier literature: 

(1) Regarding the overall sign of the pseudoscalar-pseudoscalar potential in Eq.~(\ref{pseudoscalar-pseudoscalar_potential}), we agree with the calculations of Refs.~\cite{Dob,S1}, correcting the earlier sign error in Ref.~\cite{Moody}. 

(2) The overall signs of the tensor-type potentials in Eqs.~(\ref{tensor-tensor_potential}), (\ref{pseudotensor-tensor_potential}) and (\ref{pseudotensor-pseudotensor_potential}) are opposite to those in Ref.~\cite{Dob}. 

(3) The $M^2/r$ term in $\mathcal{V}_2 + \mathcal{V}_3$ of the vector-vector potential in Eq.~(\ref{vector-vector_potential}), which arises together with a contact term, was omitted in Ref.~\cite{Dob}. 

(4) The $\mathcal{V}_3$ term (omitted in Ref.~\cite{Dob}) in the pseudovector-pseudovector potential in Eq.~(\ref{pseudovector-pseudovector_potential}) arises from a longitudinal polarisation mode for a massive spin-1 boson and nonconservation of the axial-vector current. 
Here we agree with the recent calculation of Ref.~\cite{Malta}. 
This term seems to tend to infinity in the limit of the boson mass $M \to 0$ and so it looks like the assumptions of perturbation theory are no longer justified (formally speaking, there is a violation of the perturbative unitarity bound), similarly to the violation of the perturbative unitarity bound in the high-energy scattering of longitudinal $W$ and $Z$ bosons in the standard model without a Higgs boson. 

It is instructive to consider what occurs in a renormalisable theory, such as the standard model. 
In this case, the combination of parameters $g_1^A g_2^A / M^2$ remains finite as $M \to 0$.
As an example, let us consider the case of $Z$ boson exchange between two fermions, where the $Z$ boson has purely pseudovector interactions and does not mix with the photon (i.e., $\sin(\theta_W) = 0$, where $\theta_W$ is the weak mixing angle). 
In this case, the $Z$ boson mass is given by $M = gv/2$, where $v$ is the Higgs vacuum expectation value and $g$ is the (universal) electroweak interaction constant \cite{Gordon17}. In order for the fermion masses, given by $m_f = fv/\sqrt{2}$ ($f$ is a species-dependent interaction constant), to remain finite as $M \to 0$, $v$ must also remain finite. Hence $g^2/M^2 = 4/v^2$ is independent of $M$ and remains finite as $M \to 0$. 
In such a regime, when the $\mathcal{V}_3$ term in Eq.~(\ref{pseudovector-pseudovector_potential}) gives the dominant contribution, it is appropriate to place constraints on the combination of parameters $g_1^A g_2^A / M^2$.
The relation to the renormalisability of the theory makes such a case especially interesting to study. 

In the special case of a massless vector boson, $M = 0$, Eq.~(\ref{pseudovector-pseudovector_potential}) simplifies to solely the $\mathcal{V}_2$ term. In contrast to a massive vector boson, a massless vector boson does not have a longitudinal polarisation mode, and hence there is no $\mathcal{V}_3$ term in Eq.~(\ref{pseudovector-pseudovector_potential}) for the special case $M = 0$. 

In light of the above, we believe it is worthwhile to reanalyse some earlier experiments (see, for example, \cite{Hunter2014,Leslie,Ji,RMP2017}) using the corrected potentials presented in the present paper and also to note the results of our paper for future experiments. 

Additionally, we note that it is possible to write certain potentials in a form where some of their constituent ``bare'' terms, namely with the interaction constants (which also contain particle indices) removed, are antisymmetric with respect to the permutation of particle indices $1 \leftrightarrow 2$. For example, instead of writing the scalar-pseudoscalar potential in Eq.~(\ref{scalar-pseudoscalar_potential}) in the form 
$V_{ps}(\v{r}) = C_1 \v{\sigma}_1 \cdot \hat{\v{r}} + C_2 \v{\sigma}_2 \cdot \hat{\v{r}}$, where $C_1$ and $C_2$ can be identified from Eq.~(\ref{scalar-pseudoscalar_potential}), the authors of \cite{Dob} choose to write the same expression in the form $\tilde{C}_1 \left( \v{\sigma}_1 + \v{\sigma}_2 \right) \cdot \hat{\v{r}} + \tilde{C}_2 \left( \v{\sigma}_1 - \v{\sigma}_2 \right) \cdot \hat{\v{r}}$, where $C_1 = \tilde{C}_1 + \tilde{C}_2$ and $C_2 = \tilde{C}_1 - \tilde{C}_2$. 
Here, one of the bare terms is symmetric under permutation of fermions, while the other is antisymmetric. 
The authors of \cite{Dob} note that only one combination of $\v{\sigma}_1 \pm \v{\sigma}_2$ survives for identical fermions. 
One should take this into account when searching for effects of spin-dependent forces between identical fermions. 
For instance, the authors of \cite{Hunter2014} present constraints on the antisymmetric potentials $V_7$, $V_{15}$ and $V_{16}$ for electrons, even though such potentials vanish for two identical fermions. 

\begin{table*}[t]
\caption{Properties of non-relativistic potentials induced by the exchange of spin-0 and spin-1 bosons. }  
\centeringֻֻ
\begin{tabular}{l|*{10}{c|}r|}
\diaghead(-6,1){aaaaaaaaaaaaaaaaaaaaaaaaaaaaaaaaaaaa}%
{Propertyֻֻֻֻֻֻֻֻֻֻ}{Potential}   & ~$\mathcal{V}_1$~ & ~$\mathcal{V}_2$~ & ~$\mathcal{V}_3$~ & ~$\mathcal{V}_{2} + \mathcal{V}_3$~  & $\mathcal{V}_{9,10}$  & ~$\mathcal{V}_{11}$~ & $\mathcal{V}_{12,13}$  & ~$\mathcal{V}_{14}$~ & ~$\mathcal{V}_{15}$~  \\ \hline 
Parity, $P:~(x,y,z) \to (-x,-y,-z)$ & $+$ & $+$ & $+$ & $+$  & $-$ & $-$ & $-$ & $-$ & $-$  \\ \hline 
Time-reversal symmetry, $T:~t \to -t$  & $+$ & $+$ & $+$ & $+$  & $-$ & $+$ & $+$ & $-$ & $-$  \\ \hline 
Velocity dependence  & $-$ & $-$ & $-$ & $-$  & $-$ & $-$ & $+$ & $+$ & $+$  \\ \hline 
Relevance in spectra (first-order energy shift) & $+$ & $+$ & $+$ & $+$  & $-$ & $-$ & $-$ & $-$ & $-$  \\ \hline 
Mediated by spin-0 boson & $+$ & $-$ & $+$ & $-$  & $+$ & $-$ & $-$ & $-$ & $-$  \\ \hline 
Mediated by massive spin-1 boson & $+$ & $+$ & $+$ & $+$  & $-$ & $+$ & $+$ & $-$ & $-$  \\ \hline 
Mediated by paraphoton & $-$ & $-$ & $+$ & $+$  & $+$ & $-$ & $-$ & $+$ & $+$  \\ \hline 
\end{tabular}
\label{tab1}
\end{table*}

We summarise the properties of the non-relativistic potentials induced by the exchange of spin-0 and spin-1 bosons in Table~\ref{tab1}. 
Several nuances of these potentials are also worth discussing. 
In macroscopic-scale experiments, the momentum and radial vectors appearing in Eqs.~(\ref{scalar-scalar_potential}) -- (\ref{pseudotensor-pseudotensor_potential}) may be treated as classical vectors, and symmetrised expressions such as $\{\v{p}, f(r)\}$ may be replaced by their classical value, $\{\v{p}, f(r)\} \to 2 \v{p} f(r)$. 
In atomic-scale experiments, however, the momentum and radial vectors need to be treated as quantum operators, and explicit symmetrisation in expressions such as $\{\v{p}, f(r)\}$ must be retained. 

Another difference between macroscopic-scale and atomic-scale experiments in relation to spin-dependent potentials is the manifestation of the parity- and/or time-reversal-invariance-violating nature of these potentials. 
Consider the $P,T$-violating correlation $\v{\sigma} \cdot \hat{\v{r}}$ in the term $\mathcal{V}_{9,10}$ in Eq.~(\ref{scalar-pseudoscalar_potential}). In macroscopic-scale experiments, $\hat{\v{r}}$ is a classical vector directed between macroscopically-separated bodies,
and the interaction $\v{\sigma} \cdot \hat{\v{r}}$ causes the fermion spins to precess about the vector $\hat{\v{r}}$ \cite{Ni1999,Hammond2007,Hoedl2011,Heckel2015,Ruoso2017,Rong2017}. 
In this sense, the macroscopic interaction $\v{\sigma} \cdot \hat{\v{r}}$ is reminiscent of the interaction of a fermion magnetic moment with a magnetic field. 
In atomic-scale experiments, $\hat{\v{r}}$ is a radial operator directed between atomic electrons and nucleons, and so the correlation $\v{\sigma} \cdot \hat{\v{r}}$ mixes atomic states of opposite parity and gives rise to atomic electric dipole moments \cite{Stadnik2017axion,Stadnik2018axion,Bouchiat1975EDM}. 
Specific details of how electric dipole moments are induced in atoms and molecules as a result of the $P,T$-violating potential term $\mathcal{V}_{9,10}$ can be found in Refs.~\cite{Stadnik2017axion,Stadnik2018axion}. Electric dipole moments in atoms and molecules can also be similarly induced as a result of the $P,T$-violating potential terms $\mathcal{V}_{14}$ and $\mathcal{V}_{15}$.
Both atomic- and macroscopic-scale measurements involve spin precession. 
However, the spin precession takes place about different sets of vectors. 
In macroscopic-scale experiments, spin precession takes place about the vectors $\v{B}$ and $\hat{\v{r}}$, while in atomic-scale experiments, spin precession takes place about the vectors $\v{B}$ and $\v{E}$. 



Finally, some of the potentials in Eqs.~(\ref{scalar-scalar_potential}) -- (\ref{pseudovector-pseudovector_potential}) may not be practical for numerical atomic calculations in their presented form for boson masses $M>m_1,m_2$. 
For example, in the case of a hydrogenlike system in a state with zero electron orbital angular momentum, the expectation value of the operator in Eq.~(\ref{pseudoscalar-pseudoscalar_potential}) vanishes in the limit $M \to \infty$. In this limit, we have $e^{-Mr}/r \to 4\pi \delta(\v{r})/M^2$. This $\delta(\v{r})$ term cancels the $\delta(\v{r})$ term inside the leftmost brackets in Eq.~(\ref{pseudoscalar-pseudoscalar_potential}), after integrating over the angular coordinates (likewise, the other terms in Eq.~(\ref{pseudoscalar-pseudoscalar_potential}) cancel for an arbitrary boson mass, after integrating over angular coordinates).
 Numerically, this cancellation is hard to achieve. Finite numerical precision becomes insufficient at arbitrarily large boson masses, leading to problems in numerical calculations. 
To circumvent such issues, one can instead write Eq.~(\ref{pseudoscalar-pseudoscalar_potential}) in the following equivalent form [which appears in an intermediate step of the derivation of the potential via Eq.~(\ref{FT4})]: 
\begin{equation}
\label{pseudoscalar-pseudoscalar_potential_form2}
V_{pp}(\v{r}) =  \frac{g_1^p g_2^p}{16 \pi m_1 m_2} \left( \v{\sigma}_1 \cdot \v{\nabla} \right) \left( \v{\sigma}_2 \cdot \v{\nabla} \right) \left( \frac{e^{-M r}}{r} \right)  \, ,
\end{equation}
and use integration by parts when calculating matrix elements of this operator. A similar situation occurs in Eq.~(\ref{pseudovector-pseudovector_potential}).

Note that as $M \to \infty$, matrix elements of Eq.~(\ref{pseudoscalar-pseudoscalar_potential_form2}) scale as $\propto 1/M^2$. This is a general property of the potentials in Eqs.~(\ref{scalar-scalar_potential}) -- (\ref{pseudovector-pseudovector_potential}), whose terms scale as $\propto 1/M^2$ (or faster) in the limit $M \to \infty$. One can see this property more clearly in the semi-relativistic form of the potentials presented in Appendix~\ref{AppE}.

In Eq.~(\ref{vector-vector_potential}), in addition to $\mathcal{V}_3$ of Eq.~(\ref{pseudoscalar-pseudoscalar_potential}) there is a contribution of $\mathcal{V}_2$ which is equivalent in form to:
\begin{equation}
\frac{g_1^V g_2^V}{16 \pi m_1 m_2} \v{\sigma}_1 \cdot \v{\sigma }_2
\Delta
\left( \frac{e^{-M r}}{r} \right)  \, .
\end{equation}
This expression results in two terms, by Eq.~(\ref{FT3}) in Appendix~\ref{AppC}, which cancel as $M \to \infty$ in a similar manner as above. Likewise, Eq.~(\ref{scalar-pseudoscalar_potential}) can be written as:
\begin{equation}
V_{ps}(\v{r}) =
\frac{g_1^p g_2^s}{8 \pi m_1} \left(\v{\sigma}_1 \cdot \v{\nabla} \right) \left( \frac{e^{-M r}}{r} \right)  \, ,
\end{equation}
thus highlighting its scaling $\propto 1/M^2$ as $M \to \infty$.
%
%
%

To summarise, we have derived the coordinate-space non-relativistic potentials induced by the exchange of spin-0 and spin-1 bosons, including contact terms that can play an important role in atomic-scale experiments. 
In the process, we have corrected for various errors and omissions in the earlier literature. 
These potentials are important for the interpretation of numerous experiments, including spectroscopy, torsion-pendulum, magnetometry, parity-nonconservation and electric-dipole-moment experiments, in the search for new bosons.

\section*{Acknowledgements‏}
We thank Derek Jackson Kimball, Szymon Pustelny, and Eric Adelberger for their valuable remarks.
The authors acknowledge the support by the DFG Reinhart Koselleck project, the European Research Council Dark-OsT advanced grant under project ID 695405, and the Simons and the Heising-Simons Foundations. 
F.F.~has been supported by the Polish Ministry of Science and Higher Education within the Diamond Grant (Grant No.~0143/DIA/2016/45). 
V.V.F.~was supported by the Australian Research Council (ARC) and the JGU Gutenberg Research Fellowship. 
M.G.K.~is supported by FQXi mini grant 2017-171370 and is grateful to JGU for hospitality. 
Y.V.S.~was supported by the Humboldt Research Fellowship.

\begin{appendices}
\section{Units and conventions} 
\label{AppA}
We employ the natural relativistic units $\hbar = c = 1$ and the metric signature $(+ - - -)$. 
We label space-time and spatial coordinates with Greek and Latin indices, respectively. 
We employ the Einstein summation convention for repeated indices. 
We use the following representation of the Dirac matrices: 
\begin{equation}
\label{Dirac_matrix_rep}
\gamma^0 = \left[ \begin{matrix} 1 & 0 \\ 0 & -1 \end{matrix}\right] \, , ~\gamma^i = \left[ \begin{matrix} 0 & \sigma_i \\ -\sigma_i & 0 \end{matrix}\right] \, , 
~\gamma_5 = \left[ \begin{matrix} 0 & 1 \\ 1 & 0 \end{matrix}\right] \, ,
\end{equation}
where $\sigma_i$ is the $i^\textrm{th}$ Pauli matrix.

\section{Useful identities} 
\label{AppB}
\begin{equation}
\label{id1}
[\sigma_i, \sigma_j] = 2i \varepsilon_{ijk} \sigma_k \, , 
\end{equation}
\begin{equation}
\label{id2}
\{ \sigma_i, \sigma_j \} = 2\delta_{ij} \, , 
\end{equation}
\begin{equation}
\label{id3}
\varepsilon_{ijk} \varepsilon^{imn} = \delta_j^m \delta_k^n - \delta_j^n \delta_k^m  \, , 
\end{equation}
\begin{equation}
\label{id4}
\left( \v{A} \times \v{B} \right) \cdot \left( \v{C} \times \v{D} \right) = \left( \v{A} \cdot \v{C} \right) \left( \v{B} \cdot \v{D} \right) - \left( \v{B} \cdot \v{C} \right) \left( \v{A} \cdot \v{D} \right)  \, . 
\end{equation}

\section{Fourier transforms} 
\label{AppC}
\begin{equation}
\label{FT1}
\int \frac{d^3 q}{(2 \pi)^3}  \frac{e^{i \v{q} \cdot \v{r}}}{M^2 + \left| \v{q} \right|^2 } = \frac{e^{-M r}}{4 \pi r}  \, ,
\end{equation}
\begin{equation}
\label{FT2}
\int \frac{d^3 q}{(2 \pi)^3}  \frac{ \left(\v{\sigma} \cdot \v{q} \right)  e^{i \v{q} \cdot \v{r}}}{M^2 + \left| \v{q} \right|^2 } = -i \left(\v{\sigma} \cdot \v{\nabla} \right) \left( \frac{e^{-M r}}{4 \pi r} \right)  \, ,
\end{equation}
\begin{equation}
\label{FT3}
\int \frac{d^3 q}{(2 \pi)^3}  \frac{ \left| \v{q} \right|^2  e^{i \v{q} \cdot \v{r}}}{M^2 + \left| \v{q} \right|^2 } = \left[ \delta(\v{r}) - \frac{M^2}{4 \pi r} \right] e^{-M r}  \, ,
\end{equation}
\begin{align}
\label{FT4}
\int & \frac{d^3 q}{(2 \pi)^3} \frac{ \left(\v{\sigma}_1 \cdot \v{q} \right) \left(\v{\sigma}_2 \cdot \v{q} \right)  e^{i \v{q} \cdot \v{r}}}{M^2 + \left| \v{q} \right|^2 } = \notag \\
& \frac{ \v{\sigma}_1 \cdot \v{\sigma}_2}{4 \pi} \left[ \frac{1}{r^3} + \frac{M}{r^2} + \frac{4 \pi}{3} \delta(\v{r}) \right] e^{-M r} \notag \\
& - \frac{ \left( \v{\sigma}_1 \cdot \hat{\v{r}} \right) \left( \v{\sigma}_2 \cdot \hat{\v{r}} \right) }{4 \pi} \left[ \frac{3}{r^3} + \frac{3M}{r^2} + \frac{M^2}{r} \right] e^{-M r} \, , 
\end{align}
\begin{equation}
\label{FT1a}
\int \frac{d^3 q}{(2 \pi)^3}  \frac{e^{i \v{q} \cdot \v{r}}}{ \left| \v{q} \right|^2 } = \frac{1}{4 \pi r}  \, ,
\end{equation}
\begin{equation}
\label{FT3a}
\int \frac{d^3 q}{(2 \pi)^3}   e^{i \v{q} \cdot \v{r}} =  \delta(\v{r})  \, ,
\end{equation}
\begin{equation}
\label{FT4a}
\int \frac{d^3 q}{(2 \pi)^3}  \frac{q_k q_l e^{i \v{q} \cdot \v{r}}}{ \left| \v{q} \right|^2 } = \frac{1}{4 \pi} \left[ \frac{\delta_{kl}}{r^3} - 3 \frac{r_k r_l}{r^5} + \frac{4 \pi}{3} \delta_{kl} \delta(\v{r}) \right]  \, ,
\end{equation}
\begin{equation}
\label{FT5a}
\int \frac{d^3 q}{(2 \pi)^3}  \v{q}  e^{i \v{q} \cdot \v{r}} = -i \v{\nabla} \delta(\v{r})  \, ,
\end{equation}
\begin{equation}
\label{FT6a}
\int \frac{d^3 q}{(2 \pi)^3}  \left| \v{q} \right|^2  e^{i \v{q} \cdot \v{r}} = - \Delta \delta(\v{r})  \, . 
\end{equation}

\section{Calculating coordinate-space potentials -- Three examples} 
\label{AppD}

\subsection*{1 -- Pseudoscalar-scalar potential}
Applying the Feynman rules to the tree-level process in Fig.~\ref{fig:Scattering} with vertex 1 being of the pseudoscalar type and vertex 2 being of the scalar type in Lagrangian~(\ref{Lag3}), gives the amplitude:
\begin{align}
\label{PS-S_amplitude} 
\mathcal{M}(q) = &i \left[ i^2 \bar{u}(\v{p}_{1,f}) \gamma_5 g_1^p u(\v{p}_{1,i}) \right] \left[ i \bar{u}(\v{p}_{2,f}) g_2^s u(\v{p}_{2,i}) \right] \notag \\
& \times \left[ \frac{- i}{M^2 - q^2} \right] \, , 
\end{align}
where $q = p_{1,f} - p_{1,i} = p_{2,i} - p_{2,f}$ is the 4-momentum associated with the virtual boson. 

In the non-relativistic limit, $q^2 = q_0^2 - \left| \v{q} \right|^2 \approx - \left| \v{q} \right|^2$, and the spinor products in (\ref{PS-S_amplitude}) simplify to: 
\begin{equation}
\label{PS-S_vertexA}
\bar{u}(\v{p}_{1,f}) \gamma_5 u(\v{p}_{1,i}) \approx - \v{\sigma}_1 \cdot \v{q} \, , 
\end{equation}
\begin{equation}
\label{PS-S_vertexB}
\bar{u}(\v{p}_{2,f}) u(\v{p}_{2,i}) \approx 2m_2 \, , 
\end{equation}
where $\v{p}_1 = (\v{p}_{1,i} + \v{p}_{1,f})/2$ and $\v{p}_2 = (\v{p}_{2,i} + \v{p}_{2,f})/2$ are the momenta of the two fermions, averaged over their respective initial and final states. 

The resulting non-relativistic momentum-space potential reads: 
\begin{equation}
\label{V-V_potential_p-space}
\tilde{V}(q) \approx \frac{\mathcal{M}(q)}{4 m_1 m_2} \approx \frac{i g_1^p g_2^s }{2m_1 } \frac{\v{\sigma}_1 \cdot \v{q}}{M^2 + \left| \v{q} \right|^2} \, . 
\end{equation}
The non-relativistic coordinate-space potential is related to the momentum-space potential via the three-dimensional Fourier transform: 
\begin{equation}
\label{FT_definition}
V(\v{r}) = \int \frac{d^3 q}{(2 \pi)^3}  e^{i \v{q} \cdot \v{r}} \tilde{V}(\v{q})  \, . 
\end{equation}
Using the Fourier transform (\ref{FT2}), we arrive at the coordinate-space potential in Eq.~(\ref{scalar-pseudoscalar_potential}). 

We note that, apart from the non-derivative form of the pseudoscalar interaction in (\ref{Lag3}), the derivative form of the pseudoscalar interaction is also commonly used in the literature (see, e.g., \cite{Pospelov2008axion,Stadnik2014axion,nEDM2017axion}): 
\begin{equation} 
\label{LagX_deriv}
\mathcal{L}_{\textrm{deriv.}} = - (\partial_\mu \phi)  \sum_\psi \frac{g^p_\psi }{2 m_\psi}  \bar{\psi}  \gamma^\mu \gamma_5 \psi  \, . 
\end{equation}
The form of the non-relativistic potential in Eq.~(\ref{scalar-pseudoscalar_potential}) does not depend on whether the non-derivative or derivative form of the pseudoscalar interaction is used. 
To see this explicitly, we note that, instead of the spinor product in (\ref{PS-S_vertexA}), we have the following spinor product for the derivative form of the pseudoscalar interaction: 
\begin{equation} 
\label{PS-S_vertexX_deriv}
\frac{q_\mu}{2 m_1} \bar{u}(\v{p}_{1,f}) \gamma^\mu \gamma_5 u(\v{p}_{1,i}) \approx \frac{ \v{\sigma}_1 \cdot \v{p}_1}{m_1} q_0  - \v{\sigma}_1 \cdot \v{q} \, . 
\end{equation}
In the non-relativistic limit, $\left| q_0 \right| \ll \left| \v{q} \right|$ and $\left| \v{p}_1 \right| / m_1 \ll 1$, and so the spinor product in (\ref{PS-S_vertexX_deriv}) reduces to (\ref{PS-S_vertexA}). 
Likewise, the form of the non-relativistic potential in Eq.~(\ref{pseudoscalar-pseudoscalar_potential}) also does not depend on whether the non-derivative or derivative form of the pseudoscalar interaction is used.

\subsection*{2 -- Vector-vector potential}
Applying the Feynman rules to the tree-level process in Fig.~\ref{fig:Scattering} with both vertices being of the vector type in Lagrangian~(\ref{Lag2}), gives the amplitude:
\begin{align}
\label{V-V_amplitude}
\mathcal{M}(q) = &i \left[ i \bar{u}(\v{p}_{1,f}) \gamma^\mu g_1^V u(\v{p}_{1,i}) \right] \left[ i \bar{u}(\v{p}_{2,f}) \gamma^\nu g_2^V u(\v{p}_{2,i}) \right] \notag \\
& \times \left[ \frac{i \left(g_{\mu \nu} - q_\mu q_\nu / M^2 \right)}{M^2 - q^2} \right] \, . 
\end{align}

In the non-relativistic limit, $q^2 \approx - \left| \v{q} \right|^2$, and the spinor products in (\ref{V-V_amplitude}) simplify to: 
\begin{equation}
\label{V-V_vertexA}
\bar{u}(\v{p}_{1,f}) \gamma^0 u(\v{p}_{1,i}) \approx 2m_1 \, , 
\end{equation}
\begin{equation}
\label{V-V_vertexB}
\bar{u}(\v{p}_{2,f}) \gamma^0 u(\v{p}_{2,i}) \approx 2m_2 \, , 
\end{equation}
\begin{equation}
\label{V-V_vertexC}
\bar{u}(\v{p}_{1,f}) \v{\gamma} u(\v{p}_{1,i}) \approx 2\v{p}_1 - i \v{q} \times \v{\sigma}_1 \, , 
\end{equation}
\begin{equation}
\label{V-V_vertexD}
\bar{u}(\v{p}_{2,f}) \v{\gamma} u(\v{p}_{2,i}) \approx 2\v{p}_2 + i \v{q} \times \v{\sigma}_2 \, . 
\end{equation}

The $q_\mu q_\nu / M^2$ term in the propagator does not contribute to the amplitude, because of the conservation of the vector current in both vertices. 
Retaining the leading-order spin-independent term and the leading-order spin-dependent term, yields the non-relativistic momentum-space potential: 
\begin{equation}
\label{V-V_potential_p-space}
\tilde{V}(q) \approx \frac{g_1^V g_2^V}{M^2 + \left| \v{q} \right|^2} \left[ 1 - \frac{ \left( \v{q} \times \v{\sigma}_1 \right) \cdot \left( \v{q} \times \v{\sigma}_2 \right)  }{4 m_1 m_2} \right] \, . 
\end{equation}
Performing the Fourier transform, Eq.~(\ref{FT_definition}), with the aid of the identity (\ref{id4}) and the Fourier transforms (\ref{FT1}), (\ref{FT3}) and (\ref{FT4}), we arrive at the coordinate-space potential in Eq.~(\ref{vector-vector_potential}).

\subsection*{3 -- Pseudovector-vector potential}
Applying the Feynman rules to the tree-level process in Fig.~\ref{fig:Scattering} with vertex 1 being of the pseudovector type and vertex 2 being of the vector type in Lagrangian~(\ref{Lag2}), gives the amplitude:
\begin{align}
\label{PV-V_amplitude}
\mathcal{M}(q) = &i \left[ i \bar{u}(\v{p}_{1,f}) \gamma^\mu \gamma_5 g_1^A u(\v{p}_{1,i}) \right] \left[ i \bar{u}(\v{p}_{2,f}) \gamma^\nu g_2^V u(\v{p}_{2,i}) \right] \notag \\
& \times \left[ \frac{i \left(g_{\mu \nu} - q_\mu q_\nu / M^2 \right)}{M^2 - q^2} \right] \, . 
\end{align}

In the non-relativistic limit, $q^2 \approx - \left| \v{q} \right|^2$, and the spinor products in (\ref{PV-V_amplitude}) simplify to: 
\begin{equation}
\label{PV-V_vertexA}
\bar{u}(\v{p}_{1,f}) \gamma^0 \gamma_5 u(\v{p}_{1,i}) \approx 2 \v{\sigma}_1 \cdot \v{p}_1 \, , 
\end{equation}
\begin{equation}
\label{PV-V_vertexB}
\bar{u}(\v{p}_{2,f}) \gamma^0 u(\v{p}_{2,i}) \approx 2m_2 \, , 
\end{equation}
\begin{equation}
\label{PV-V_vertexC}
\bar{u}(\v{p}_{1,f}) \v{\gamma} \gamma_5 u(\v{p}_{1,i}) \approx 2m_1 \v{\sigma}_1 \, , 
\end{equation}
\begin{equation}
\label{PV-V_vertexD}
\bar{u}(\v{p}_{2,f}) \v{\gamma} u(\v{p}_{2,i}) \approx 2\v{p}_2 + i \v{q} \times \v{\sigma}_2 \, . 
\end{equation}

Again, the $q_\mu q_\nu / M^2$ term in the propagator does not contribute to the amplitude, because of the conservation of the vector current in the second vertex. 
We hence find the following non-relativistic momentum-space potential: 
\begin{equation}
\label{PV-V_potential_p-space}
\tilde{V}(q) \approx \frac{g_1^A g_2^V }{M^2 + \left| \v{q} \right|^2} \left[ \v{\sigma}_1 \cdot \left( \frac{ \v{p}_1}{m_1} - \frac{\v{p}_2}{m_2} \right) + i \frac{\left( \v{\sigma}_1 \times \v{\sigma}_2 \right) \cdot \v{q}}{2 m_2}  \right] \, . 
\end{equation}
Performing the Fourier transform, Eq.~(\ref{FT_definition}), with the aid of the Fourier transforms (\ref{FT1}) and (\ref{FT2}), and then performing the symmetrisation $\left( \v{p}_1/m_1 - \v{p}_2/m_2 \right)  \left( e^{-M r}/r \right)  \to \frac{1}{2} \{ \v{p}_1/m_1 - \v{p}_2/m_2 , e^{-M r}/r \} $, we arrive at the coordinate-space potential in Eq.~(\ref{pseudovector-vector_potential}).

\section{Coordinate-space potentials in semi-relativistic form} 
\label{AppE}
In a form convenient for numerical atomic calculations using relativistic Dirac-Hartree-Fock wavefunctions, the potentials in Eqs.~(\ref{scalar-scalar_potential}) -- (\ref{pseudovector-pseudovector_potential}), including Dirac spinors, can be written in the following general form: 
\begin{equation}
\label{S-S_rel}
V_{ss}(\v{r}) = -
\left( \bar{\psi}_2   g^s_2   \psi_2 \right)
\left( \bar{\psi}_1   g^s_1   \psi_1 \right)
\frac{e^{-M r}}{4 \pi r}
 \, ,
\end{equation}

\begin{equation}
V_{ps}(\v{r}) = -
\left( \bar{\psi}_2   g^s_2   \psi_2 \right)
\left( \bar{\psi}_1  i \gamma_{5} g^p_1   \psi_1 \right)
\frac{e^{-M r}}{4 \pi r}
  \, ,
\end{equation}

\begin{equation}
V_{pp}(\v{r}) = -
\left( \bar{\psi}_2   i \gamma_{5} g^p_2   \psi_2 \right)
\left( \bar{\psi}_1  i \gamma_{5} g^p_1   \psi_1 \right)
\frac{e^{-M r}}{4 \pi r}
  \, ,
\end{equation}

\begin{equation}
V_{VV}(\v{r}) =
\left( \bar{\psi}_2   \gamma^{\mu} g^V_2   \psi_2 \right)
\left( \bar{\psi}_1  \gamma_{\mu} g^V_1   \psi_1 \right)
\frac{e^{-M r}}{4 \pi r}
  \, ,
\end{equation}

\begin{equation}
V_{AV}(\v{r}) =
\left( \bar{\psi}_2   \gamma^{\mu} g^V_2   \psi_2 \right)
\left( \bar{\psi}_1  \gamma_{\mu} \gamma_5 g^A_1   \psi_1 \right) 
\frac{e^{-M r}}{4 \pi r}
 \, ,
\end{equation}

\begin{equation}
\label{PV-PV_rel}
V_{AA}(\v{r}) =
\left( \bar{\psi}_2   \gamma^{\mu} \gamma_5 g^A_2   \psi_2 \right)
\left( \bar{\psi}_1  \gamma_{\mu} \gamma_5 g^A_1   \psi_1 \right) 
\frac{e^{-M r}}{4 \pi r}
 \, ,
\end{equation}
where we have made use of the static approximation for the boson propagators, and in Eq.~(\ref{PV-PV_rel}), we have dropped additional terms arising from the longitudinal polarisation mode of the massive spin-1 boson. 
In practical applications, often one of the fermions can be treated non-relativistically, in which case the potentials in Eqs.~(\ref{S-S_rel}) -- (\ref{PV-PV_rel}) reduce to a mixed relativistic/non-relativistic form (see, e.g., Refs.~\cite{Stadnik2017vector,Stadnik2017axion,Stadnik2018axion} for more details). 


\end{appendices}

\bibliographystyle{prsty}

\end{document}